# Symmetry-mediated quantum coherence of $W^{5+}$ spins in an oxygen-deficient double perovskite

Shannon Bernier,[a] Mekhola Sinha,[a] Tyler J. Pearson,[b] Peter V. Sushko,[c] Paul H. Oyala,[d] Maxime A. Siegler,[a] W. Adam Phelan,[a,e,‡], Abby N. Neill,[a] Danna E. Freedman,[b,%], and Tyrel M. McQueen[a,e,f,#]

[a] Department of Chemistry, The Johns Hopkins University, Baltimore, MD 21218, USA

[b] Department of Chemistry, Northwestern University, Evanston, IL 60208, USA

[c] Physical and Computational Sciences Directorate, Pacific Northwest National Laboratory, 902 Battelle Boulevard, Richland, WA 99354, USA

[d] Division of Chemistry and Chemical Engineering, California Institute of Technology, Pasadena, CA 91125, USA

‡Present Address: Los Alamos National Laboratory, NM 87544, USA

[e] Institute for Quantum Matter, William H. Miller III Department of Physics and Astronomy, The Johns Hopkins University, Baltimore, MD 21218, USA

[f] Department of Materials Science and Engineering, The Johns Hopkins University, Baltimore, MD 21218, USA

% Present Address: Department of Chemistry, Massachusetts Institute of Technology, Cambridge, MA 02139, USA

([#] corresponding author: mcqueen@jhu.edu)

## Abstract:

Elucidating the factors limiting quantum coherence in real materials is essential to the development of quantum technologies. Here we report a strategic approach to determine the effect of lattice dynamics on spin coherence lifetimes using oxygen deficient double perovskites as host materials. In addition to obtaining millisecond $T_1$ spin-lattice lifetimes at T ~ 10 K, measurable quantum superpositions were observed up to room temperature. We determine that $T_2$ enhancement in $Sr_2CaWO_{6-\delta}$ over previously studied $Ba_2CaWO_{6-\delta}$ is caused by a dynamically-driven increase in effective site symmetry around the dominant paramagnetic site, assigned as $W^{5+}$ *via* electron paramagnetic resonance spectroscopy. Further, a combination of experimental and computational techniques enabled quantification of the relative strength of spin-phonon coupling of each phonon mode. This analysis demonstrates the effect of thermodynamics and site symmetry on the spin lifetimes of $W^{5+}$ paramagnetic defects, an important step in the process of reducing decoherence to produce longer-lived qubits.

## Introduction:

Quantum information science represents the next frontier in myriad fields including sensing, computing, and metrology.[1-3] Advances in this arena are enabled by new developments in material hosts for the smallest unit of quantum information - the quantum bit or qubit - each tailored for a specific use.[4-5] Many quantum computing architectures have been explored to date by materials scientists, including topological,[6-7] trapped ion,[8-9] superconducting,[10] molecular spin-based,[11] photonic,[12-13] and defect-based.[14] For the design of quantum sensors, defect-based electronic spins housed in nuclear spin-free matrices possess broad appeal for their *in situ* optical addresS1lity and relatively easy-to-characterize phonon structures.[15-16] Electronic spins in general are also appealing for this application because their relaxation properties are extremely sensitive to their chemical environment and tunable using well-established chemical principles.[17] Dramatic advances in exerting control over electronic spin-based qubits have been made in the last two decades.[18-22]

The viability of a qubit is a function of two related but distinct parameters: $T_1$, related to the spin lattice relaxation time, and $T_2$, related to the spin-spin relaxation time.[23] Spin-lattice relaxation is dictated by the interaction between the electronic spin and its local and extended phonon bath. Spin-spin relaxation is primarily a function of the interaction between an individual spin and its neighboring spin centers and magnetic fields. $T_2$ is also a measure of the amount of time that coherent information can be recovered from an excited system, and for this reason, it has been the focus of studies across potential qubit hosts. It has been extensively demonstrated that the best way to lengthen $T_2$ at liquid helium temperatures is through the rigorous exclusion of nuclear spins combined with dilution of electronic spins; both approaches work to prevent unwanted interactions.[21,24-28] At the same time, $T_2$ has a physical maximum: $T_2 \leq 2\times T_1$. Thus, at higher temperatures lengthening $T_2$ will necessarily require doing the same to $T_1$.

The factors that control $T_1$ in solid state systems are experimentally understudied relative the factors influencing $T_2$.[29-33] The development of design principles for lengthening $T_1$ has consequently become an extremely active area of research.[34-35] In this work we have prioritized understanding the specific chemical and structural principles which dictate $T_1$ and $T_2$ in our qubit host materials. $T_1$ is much more strongly influenced by lattice vibrations to which $T_2$ is largely insensitive,[36] but not immune, as will be shown in this work. At its core, optimization of $T_1$ requires careful engineering of the vibrational and phonon modes of the spin host material and of the interaction between the spin and its surrounding lattice. Such strategies as stiffening the vibrational modes surrounding the spin site and designing materials out of only lighter elements (with inherently less spin-orbit coupling) have proven encouraging for maximizing $T_1$, but more general design principles remain under investigation.[30,37]

To more fully understand the ways that localized electronic spins interact with the phonon bath of their host matrix, we have chosen oxygen deficient double perovskites as a model system. Double perovskites are a well-studied class of materials and have been explored in a wide range of quantum materials studies.[38-40] The wide band-gap of these materials helps to energetically isolate defect sites above cryogenic temperatures.[41] Previously, we reported the discovery of extraordinarily long $T_1$ times of paramagnetic sites embedded in the oxygen deficient double perovskite $Ba_2CaWO_{6-\delta}$ and hypothesized that the origin of the paramagnetic spin was $W^{5+}$.[29] Using a combination of heat capacity measurements and pulse electron paramagnetic resonance (EPR) spectroscopy, the coherence properties of these defect sites were explained as a sum of Raman processes, a finding consistent with recent reports.[11,42-45] Furthermore, by calculating the phonon modes' relative contributions to the relaxation dynamics in that work, we were able to derive new insight into which modes promote decoherence. Specifically, we determined that the vibrations of lighter elements contributed more to decoherence than those of heavier elements. This conclusion has since been further supported by numerous studies and has become a design principle in constructing solid-state electronic spin-based qubits.[46-48]

Building on these findings, here we set out to expand the study of structure-property correlations to other systems, starting with $Sr_2CaWO_{6-\delta}$, a lower mass chemical analogue to $Ba_2CaWO_{6-\delta}$. We hoped that the smaller spin-orbit coupling present in the strontium analogue would render the spin centers less sensitive to motion of the lattice and thereby prolong relaxation times. We were further encouraged by the fact that Sr, like Ba, has a low natural abundance of nuclear spin-active nuclei.

In this system we indeed observed coherence times measurable up to room temperature, a substantial improvement over the $Ba_2CaWO_{6-\delta}$ system and a validation of our initial hypothesis. Interestingly, we also noted an unusual elongation of coherence time $T_2$ with increasing temperature. This prompted us to identify the precise ion involved in the paramagnetic defect centers, an unanswered question from our previous work. To perform these characterizations, we employed a combination of one-and two-dimensional pulse EPR, hyperfine sublevel correlation (HYSCORE) spectroscopy, heat capacity measurements, thermogravimetric analysis (TGA), powder x-ray diffraction (PXRD), single crystal x-ray diffraction (SCXRD), magnetization, attenuated total reflectance infrared spectroscopy (ATR-IR), crystal field calculations, and density functional theory (DFT). Using this suite of techniques, we show that local distortions in the environment of the paramagnetic site contribute to an unusual lengthening in $T_2$ with increasing temperature. Finally, we validate our previous finding that double perovskite qubit hosts can house long-lived electronic spin-based qubits with measurable superposition lifetimes out to room temperature. This represents substantial progress towards using these materials in quantum sensing devices.

## Results:

Similar to the previously reported Ba analogue, $Sr_2CaWO_{6-\delta}$ is an $A_2BB'O_6$ double perovskite consisting of alternate corner-sharing octahedra of B = $CaO_6$ and B' = $WO_6$. Sr cations fill the A sites of the framework. Rietveld refinement to the PXRD data (Figure S2, Table S3) shows that the material adopts a monoclinic structure ($P2_1/n$, space group #14) at room temperature. See SI §S2 for additional discussion. These data are supported by SCXRD results at T = 213 K (data presented in SI §S3). Pawley fits (Figure S3, Table S4) to PXRD data also indicate that the material undergoes no structural phase transitions from T = 12 K to room temperature (SI §S4). Geometric constraints in this simple unit cell imply that miniscule variations in lattice parameters are due to the tilting of the $CaO_6$/$WO_6$ octahedra. Such changes have been shown to affect paramagnetic sites in other solid-state materials.[49]

Oxygen vacancies produce $W^{5+}$ sites on which single electronic spins are localized (Figure 1a) as supported by spin density estimates from HYSCORE spectroscopy (data in SI §S5-S6). This technique provides a direct measure of the strength of the interaction between unpaired electronic spins and nuclear-spin active nuclei. As will be detailed below, the degree of spin localization may also be determined as a function of temperature. In order to quantify the precise quantity of oxygen vacancies, we first turned to TGA measurements combined with ATR-IR (data in SI §S7-S8) but, due to sample aging (SI §S9) analysis of these data was unsuccessful. Instead, an oxygen deficiency of about 1.18% ($\delta \approx 0.07$) was identified *via* magnetization measurements using the protocol described in the Methods section, data presented in SI §S10. Measurements on a range of controllably reduced $Sr_2CaWO_{6-\delta}$ polycrystalline powders (SI §S11) support this determination using magnetization alone. Magnetization data were also used to exclude the possibility of a magnetic phase in this material (Figures S24 and S25).

**Spin-spin relaxation – EPR and tungsten site symmetry:**

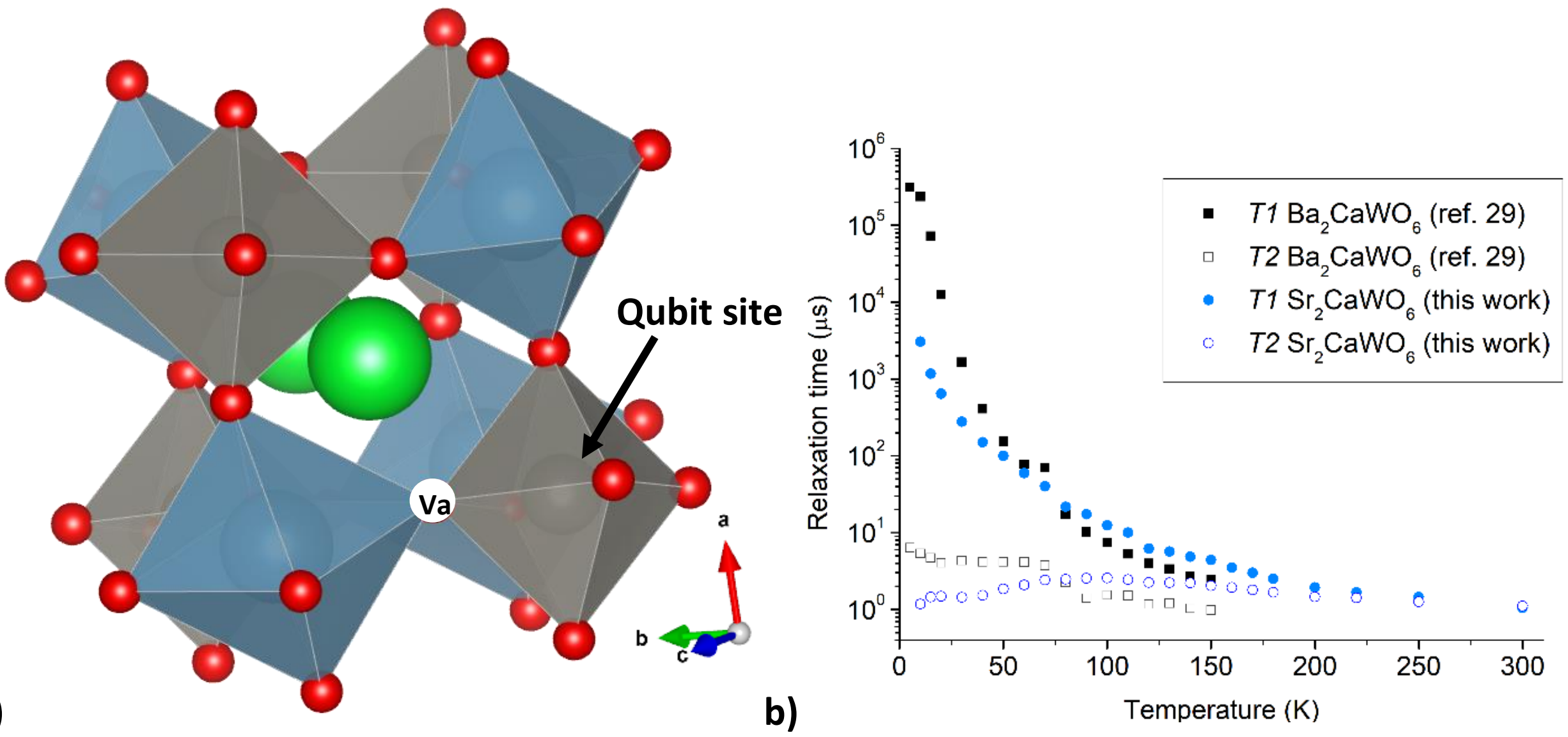

**a)** **b)**

**Figure 1: Relaxation time of paramagnetic electron at tungsten site.** a) Partial unit cell for $Sr_2CaWO_6$ illustrating the qubit site located next to a proposed oxygen vacancy labelled Va. Red = oxygen, gray = tungsten, blue = calcium, green = strontium. b) $T_1$ and $T_2$ relative to temperature for the sharp, long-lived feature at approximately 3500 G for both $Sr_2CaWO_6$ and its chemical analogue $Ba_2CaWO_6$.

To determine the viability of this system as a host matrix for defect-based qubits, we measured $T_1$ and $T_2$ from T = 5 to 300 K using pulse EPR. Figure 1b depicts relaxation times for the dominant feature observed at $B_0$ = 3500 G in both $Sr_2CaWO_6$ and the previously studied analogue compound $Ba_2CaWO_6$. In both materials, we observe decreasing $T_1$ with increasing temperature. As previously reported, $T_2$ for the barium compound also decreases as temperature increases.[29] In the strontium compound, we initially see a small rise in $T_2$ at low temperature. This increase may be explained as a reducing contribution of spectral diffusion to relaxation, common in other materials below T = 10 K. Much less commonly observed, however, is the doubling of $T_2$ between T = 30 K and 130 K that we encounter in $Sr_2CaWO_6$. $T_2$ is, in theory, a metric for a temperature-independent process, depending only on the spin-spin interactions within the system and an upper limit set by $T_1$. Here, $T_1$ decreases over the range where $T_2$ increases, suggesting that the peculiar increase in coherence time is not driven by spin-lattice dynamics nor an increase in the maximum possible value for $T_2$.

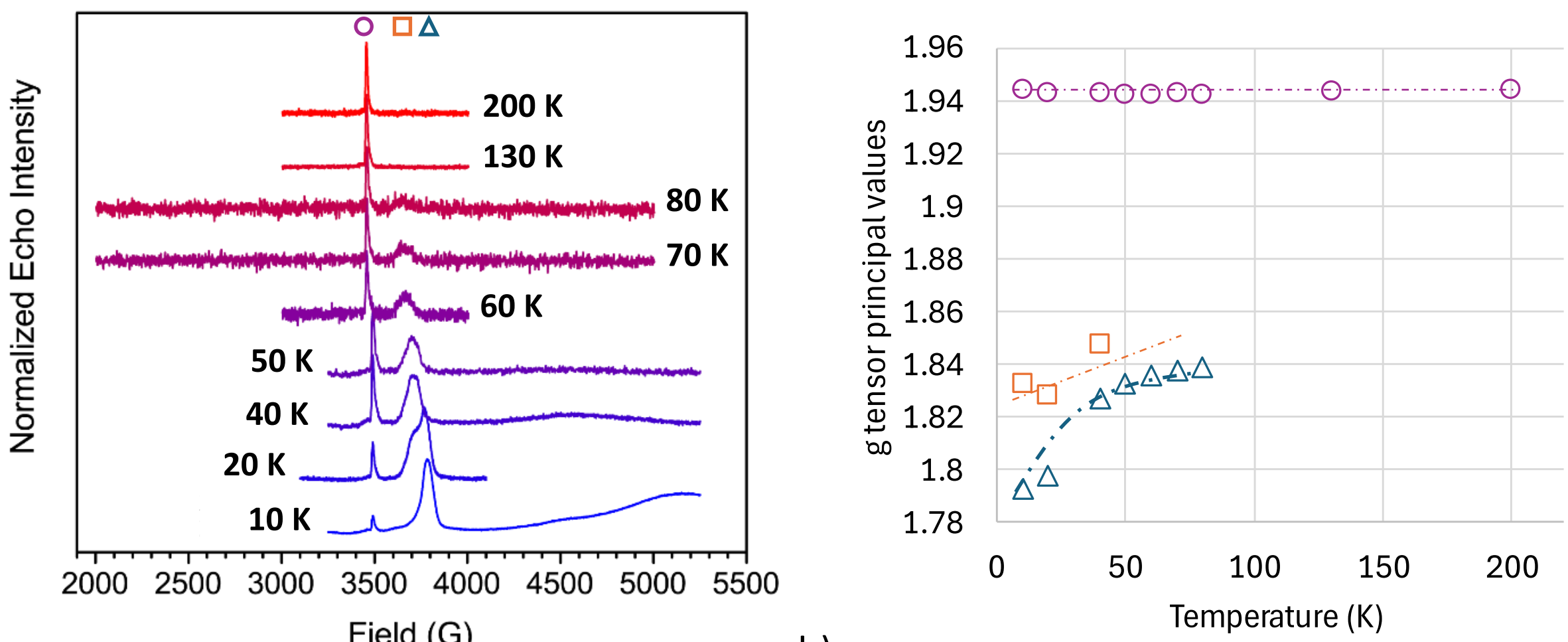

a) b)

**Figure 2: EDFS EPR spectra and calculated g tensors.** a) EDFS EPR spectra of a powder of $Sr_2CaWO_{6-\delta}$. b) g tensors produced by fitting data in a) to a model of 3 symmetric tensors, with lines to guide the eye. Above 40 K, only two g tensors were required to capture peak shape; above 100 K, only one was required.

Echo-detected field swept (EDFS) EPR spectra (Figure 2a) of $Sr_2CaWO_6$ show two features at low temperature. This is in contrast to our previous observations for the chemical analogue $Ba_2CaWO_6$, which showed a single feature with g = 1.96 at all temperatures. [29] The strength of these features quantifies the electron-ligand field coupling and thus local distortions and Einstein modes in the material. An anisotropic feature is initially dominant at low temperature. Between T = 10-130 K the strength of this anisotropic feature is reduced as an isotropic feature at 3500 G (g ≅ 1.943) becomes dominant (see also Figure S32). The increasing dominance of the isotropic signal corresponds roughly with the temperature region where we observe the $T_2$ increase. Therefore, we posit that higher isotropy in the EPR data results in higher measured $T_2$ values even while limited by decreasing $T_1$.

As $W^{4+}$ and $W^{6+}$ are not EPR active, we assign the tungsten valency as $W^{5+}$. The spin density estimates obtained from isotropic hyperfine coupling of $^{183}W$ species in the material show that the dominant phase exhibiting the signal at $g = 1.99$ in the HYSCORE data (above T ~ 70 K; present as the isotropic component in Figure 2a) corresponds to electrons localized mostly on oxygen orbitals at these sites, with at least some spin localization directly at the tungsten site. The formal oxidation state of reduced tungsten atoms in the material is $W^{5+}$ with strong oxygen covalency for all temperatures measured. This assignment is detailed in Table S8 and the increasing isotropy of the HYSCORE data with increasing temperature may be observed in Figures S6-S10. This increase signifies that either the spin density is becoming more spherically symmetric – such as spin localizing in an s-orbital, or more evenly distributing between a set of d-orbitals – or that the through-space dipolar interaction of the $^{183}W$ nucleus with spin density localized on adjacent atoms is becoming weaker.

In the absence of evidence to support or deny the second hypothesis, we instead consider the idea that spin density (from an unpaired electron) is becoming more spherically symmetric as temperature increases. Spin density localizing in s-orbitals is unlikely for a 5*d* transition metal such as tungsten; $W^{5+}$ is expected to be $5d^1$. Therefore, the increasing isotropy of the HYSCORE data indicates that these $5d^1$ paramagnetic electrons are driving the increase in symmetry. We refer to this phenomenon as an increase in "effective" site symmetry because such spin delocalization across d-orbitals would not change the physical symmetry of the site in the lattice but would result in a more symmetrical electronic environment at that site. In the monoclinic lattice, the 5$d^1$ unpaired electron occupies a single $d_{z^2}$ orbital with preferred orientation along the axis connecting the two closest strontium atoms (the body diagonal of the unit cell). Lattice vibrations in this geometry result in a variation of strontium position, leading in turn to a change in the preferred $d_{z^2}$ orbital orientation. When averaged over many such vibrations, this results in a nearly spherical spin density. See Figure S34 and SI §S13 for additional discussion.

Note that these unpaired electrons are introduced to the tungsten sites by the loss of an oxygen atom from a formerly symmetric full octahedron. Such reduction of $Sr_2CaWO_6$ necessarily requires charge balancing in the form of changing oxidation states for other ions in the lattice - either additional tungsten atoms or a neighboring calcium or strontium. For simplicity, we will limit the following discussion to the tungsten atoms only, as $Ca^{1+}$ or $Sr^{1+}$ are chemically dubious and spin localization on these ions is not supported by our DFT results, discussed below. Rather, DFT results suggest one spin localizing on a specific tungsten atom with the other spin delocalized across other tungsten atoms.

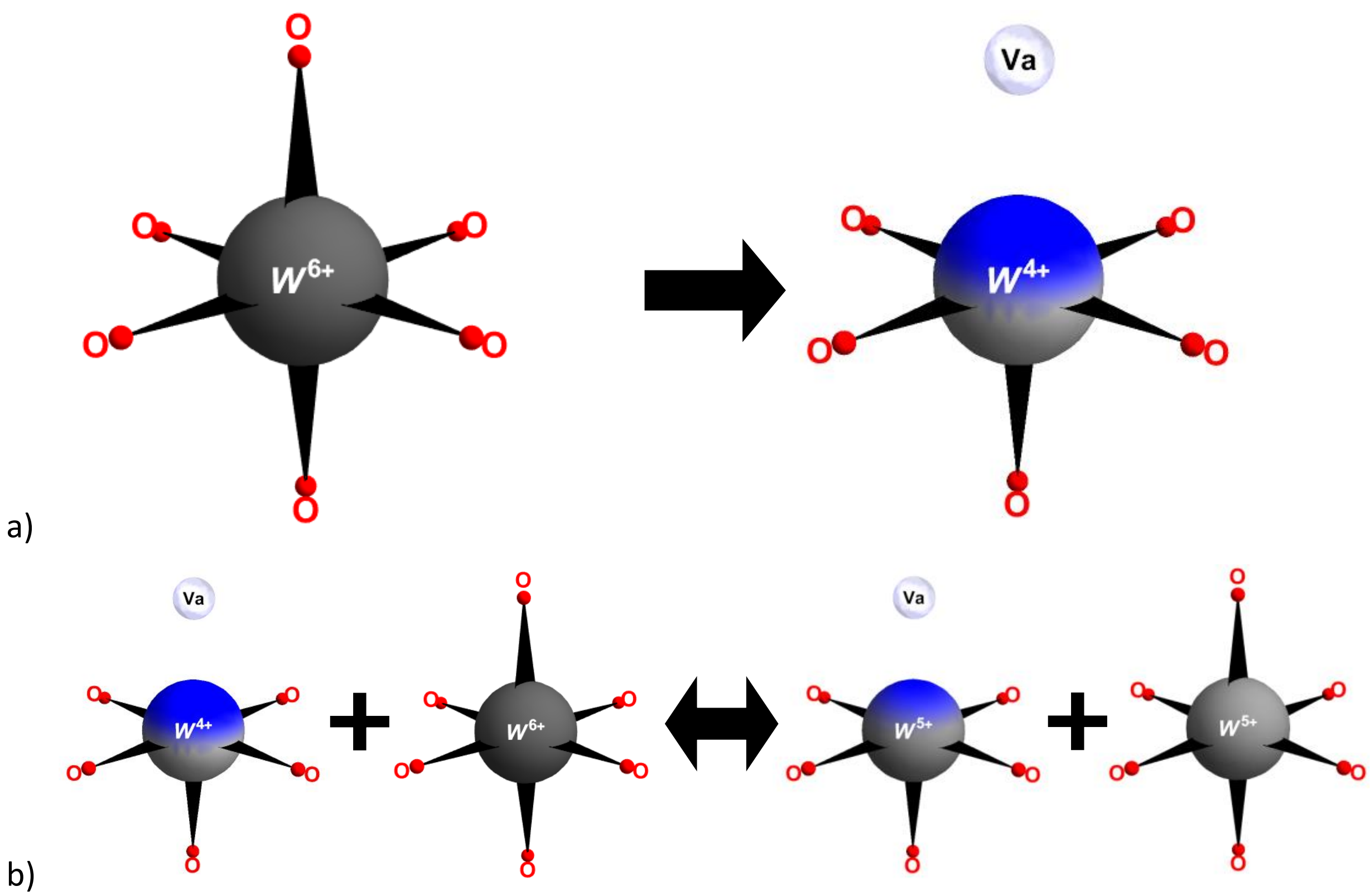

**Figure 3: Schematic of reduction of a tungsten site.** a) Process by which a $W^{6+}$ site of the form present in fully oxidized $Sr_2CaWO_6$ is converted into a $W^{5+}$ site with asymmetric ligand field by the loss of an oxygen atom. b) Equilibrium between a pairing of $W^{4+}$ and $W^{6+}$ sites with two $W^{5+}$ sites, one with an asymmetric and one with a symmetric ligand field.

In the fully oxidized material, all tungsten atoms are $W^{6+}$ ions. $W^{6+}$ is not EPR-active, so both full and incomplete octahedra are "silent" in our data. As shown in Figure 3, the loss of a single oxygen bound to a tungsten site would yield a $W^{4+}$ ion with an asymmetric ligand field. The resulting site could produce an anisotropic EPR signal but is not supported by our HYSCORE data as the origin of the detected signal. Instead, one of the extra electrons on a $W^{4+}$ site hops onto a nearby $W^{6+}$ with a complete octahedron, producing one $W^{5+}$ site with an asymmetric ligand field and another $W^{5+}$ site with a symmetric ligand field. We assign these two sites as responsible for the anisotropic and isotropic signals we detect, respectively. DFT estimates put the energy of this electron transfer at 0.1-0.3 eV. We find that the resulting vacancy is a shallow electron donor, meaning this process is likely to take place at low temperatures.

An additional, complementary effect is oxygen vacancy diffusion in the vicinity of an asymmetric $W^{5+}$ center. An oxygen vacancy may occur at any of the six oxygen atoms coordinated to a central tungsten. As supported by our DFT results (Figure S37), the diffusion of this vacancy from one position to another also serves to qualitatively increase the symmetry of the electronic environment around the central tungsten. Notably, the spin density maxima produced by averaging over $d_{z^2}$ orbital orientation align with the minima of spin density produced by averaging over different vacancy locations in the unit cell. The result is that, when taken together, these effects combine to produce higher effective $W^{5+}$ site symmetry even for partial oxygen octahedra as depicted in Figure 4.

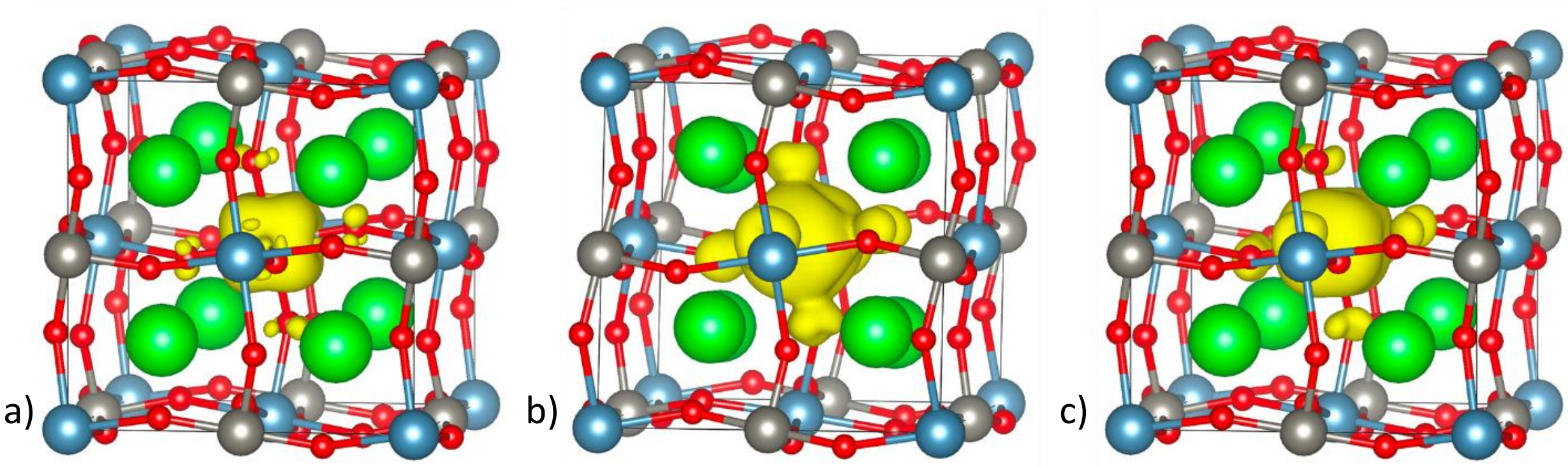

**Figure 4: Qualitative illustration of increasing spin symmetry.** Green = calcium, blue= strontium, gray = tungsten, red = oxygen, yellow = spin density of $5d^1$ electron on tungsten. To produce this qualitative image, weightings were assumed to be equal for each possible configuration. a) Average spin density of four different orientations of $d_{z^2}$ orbitals along each unit cell body diagonal, localized on central tungsten atom. b) Average spin density of six possible locations for oxygen vacancy around central tungsten atom. c) Average spin density of four different orientations of $d_{z^2}$ orbitals and six possible locations for oxygen vacancy, illustrating the synergy of the two effects in producing a highly-symmetrical (more spherical) spin density.

Vacancy diffusion becomes more frequent for partial octahedra as temperature increases, producing a more spherical distribution of spin on the tungsten site. Furthermore, the ligand field around an asymmetric $W^{5+}$ will also experience increased thermal motion, leading to a more symmetric electronic environment for this spin. At the same time, thermal motion yields greater $d_{z^2}$ orbital averaging for both partial and complete octahedra. The population of paramagnetic electrons in partial octahedra also decreases as these electrons are thermally activated to hop onto sites with complete octahedra. Since $T_1$ and $T_2$ were measured at 3500 G, the increase in isotropic signal at 3500 G may be directly related to increasing population of complete octahedra. The end result of any of these effects is the same: an increase in the total number of paramagnetic electrons in effective high-symmetry ligand fields. Once the population of $W^{5+}$ in complete octahedra is saturated, the normal effects which serve to decrease coherence time start to dominate over vacancy diffusion and thermal motion, ultimately leading to a decrease in relaxation time as the material approaches room temperature.

**Spin-lattice relaxation - Heat capacity and phonon modes:**

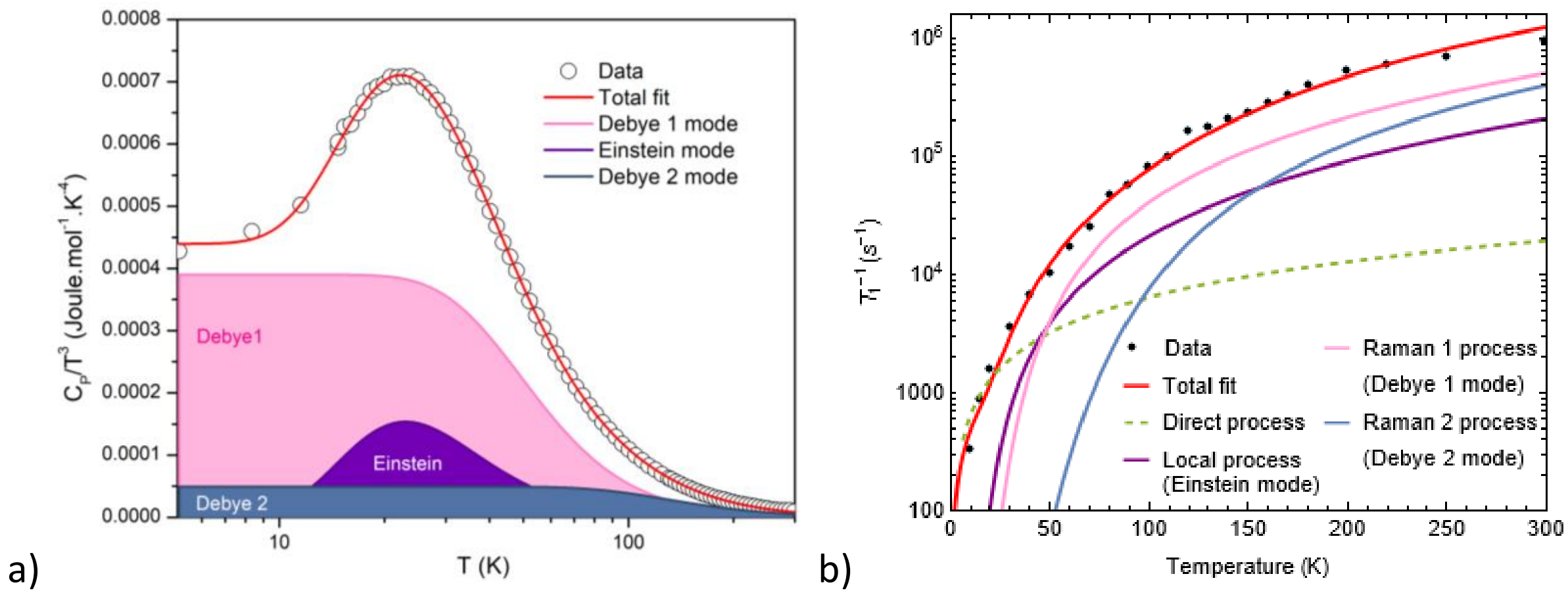

**Figure 5: Heat capacity and EPR fits.** a) Heat capacity ($C_p$) divided by temperature cubed ($T^3$) versus temperature (T) plotted on a logarithmic scale. The red line shows a fit to the experimental data. Contributions of individual components are plotted below: pink – Debye 1 phonon mode, blue – Debye 2 phonon mode, purple – low energy Einstein phonon mode. b) Inverse of longitudinal relaxation time ($T_1$) relative to temperature obtained from pulse EPR. The red line shows the fit to obtain the contributions of each of the decoherence processes (direct and local process corresponding to the Einstein mode and the Raman processes corresponding to the Debye modes). The characteristic temperature for each mode was fixed to those obtained from the heat capacity fit.

To identify the mechanisms restricting $T_1$ in this system, we next turned to a combination of heat capacity and EPR measurements. When plotted as $Cp/T^3$ vs. T, heat capacity data collected *via* the protocol detailed in the Methods section roughly approximate the 1-D phonon density of states (DOS). As shown in Figure 5a, the data for $Sr_2CaWO_{6-\delta}$ are best described by a combination of three modes: two acoustic Debye phonon modes and one low-lying Einstein optical mode. At sufficiently low temperatures Debye modes plateau at a constant value, while an Einstein mode accounts for peaking behavior.[50] The oscillator strength (Table 1) adds up to ~ 10.3(5): a confirmation that our model accurately accounts for the 10 atoms per formula unit. No evidence of phase transition has been found in the T = 2-300 K range.

The presence of an Einstein mode is indicated by a peak at T ≅ 22 K in Figure 5a. We ascribe the origin of this Einstein mode to activated local vibrational processes related to the tilting of $WO_6$/$CaO_6$ octahedral units (Figure S33), a common structural distortion in perovskites. The appearance of such a low-lying Einstein mode in this system is surprising but not unprecedented in double perovskites.[39,51] Indeed, we observed similar features in our earlier analysis of the barium analogue.[29] The Einstein mode in $Sr_2CaWO_6$ occurs at a nearly identical temperature but is much weaker than that of $Ba_2CaWO_6$, which results in a lower overall heat capacity in the strontium analogue.

Additional insight into the types and effects of phonon modes is garnered by comparison to our EPR data. Figure 5b shows a plot of the inverse of $T_1$ with respect to temperature generated from our pulse EPR data. These data were fit to a model described in the Methods section following the process pioneered in our earlier work on the barium analogue,[29] where $A_{dir}$, $A_{ram1}$, $A_{ram2}$, and $A_{loc}$ are coefficients representing the strength of one direct, two Raman, and one local process, respectively. Each Debye mode (corresponding to a Raman process) is associated with a characteristic Debye temperature ($\theta_{D1}$, $\theta_{D2}$) and the local mode with an Einstein temperature ($\theta_E$) which are fixed to values obtained from the quantitative analysis of the heat capacity (Table 1).

**Table 1:** Characteristic temperatures and number of oscillators used in fitting Einstein and Debye modes to the specific heat for $Sr_2CaWO_{6-\delta}$.

| Mode | Temperature (K) | Oscillator strength per formula unit | Relative spin-phonon coupling |
|---|---|---|---|
| Einstein | $\theta_E$ = 113.3(4) | 0.75(1) | 2.12 |
| Debye 1 | $\theta_{D1}$ = 242(2) | 2.8(1) | 1 |
| Debye 2 | $\theta_{D2}$ = 641(29) | 6.7(3) | 6.04 |

Pulse EPR data below T = 10-15 K are well-modeled with the single-phonon direct process, as is typical.[29] This process is best described as a spin flip because the faster phonon-mediated processes are minimally activated at this temperature. Where T << $\theta_D$, this process contributes to a linear increase in $T_1^{-1}$ with temperature. The orders-of-magnitude greater influence of the direct process in $Sr_2CaWO_6$ compared to $Ba_2CaWO_6$ (Table 2) provides an explanation for the lower $T_1$ values in this material at low temperature.[29]

**Table 2:** Coupling constants used in fitting $1/T_1$ data for $Sr_2CaWO_{6-\delta}$ and $Ba_2CaWO_{6-\delta}$.

| Process | Fit parameter for $Sr_2CaWO_{6-\delta}$ (this work) | Fit parameter for $Ba_2CaWO_{6-\delta}$ (Ref. 29) |
|---|---|---|
| Direct | $A_{dir}$ = 64(6) $K^{-1}s^{-1}$ | $A_{dir}$ = 0.49(09) $K^{-1}s^{-1}$ |
| Raman 1 | $A_{ram1}$ = 2.4(9)·$10^6$ $s^{-1}$ | $A_{ram1}$ = 4.4(8)·$10^6$ $s^{-1}$ |
| Raman 2 | $A_{ram2}$ = 1.7(6)·$10^7$ $s^{-1}$ | $A_{ram2}$ = 3.2(8)·$10^8$ $s^{-1}$ |
| Local | $A_{loc}$ = 3(1)·$10^4$ $s^{-1}$ | $A_{loc}$ = 8(2)·$10^3$ $s^{-1}$ |

In the intermediate temperature regime above 15 K, the data are better modeled using the Raman process, wherein two different phonons mediate spin-lattice relaxation. Contributions from the direct process and the local Einstein mode have a minor influence for $Sr_2CaWO_6$ in the low temperature (T < 30 K) regime, but the Raman processes dominate relaxation across almost the entire temperature range. This runs counter to the model often invoked in the fitting of the temperature dependence of $T_1$, wherein the Raman process is expected to give way to localized phonon modes at the highest temperature extreme. Finally, quasi-local phonons provide a small contribution at temperatures above T = 70 K, analogous to local vibrations in the immediate vicinity of the spin center.

We also explored the relative influence on relaxation as a whole. The fit parameters obtained from the heat capacity and the spin relaxation lifetime data can be combined to gain quantitative information about the relative strength of spin-phonon coupling of each of the modes using the following equations:

$$\frac{A_{ram2}}{A_{ram1}} = \left(\frac{\mathrm{N_{D2}}}{\mathrm{N_{D1}}}\right)^2 \left(\frac{\theta_{\mathrm{D1}}}{\theta_{\mathrm{D2}}}\right) \left(\frac{G_{ram2}}{G_{ram1}}\right)^4 \tag{1}$$

$$\frac{A_{loc}}{A_{ram}} = \frac{36\,\pi^5}{9 \cdot 6^{\frac{10}{3}}\,\pi^{\frac{11}{3}}} \left(\frac{\mathrm{N_E}}{\mathrm{N_D}}\right)^2 \left(\frac{\theta_{\mathrm{D}}}{\theta_{\mathrm{E}}}\right) \left(\frac{G_{loc}}{G_{ram}}\right)^4 \tag{2}$$

where N is the oscillator strength of each of the modes, θ is the characteristic temperature and G is a parameter describing coupling between lattice vibrations and the spin of the paramagnetic electron.[52] Values are reported in Table 1. From this analysis we observe that the spin-phonon coupling parameter is strongest for the $\theta_{D2}$ = 641(29) K mode and weakest for the $\theta_{D1}$ = 242(2) K mode. This behavior is qualitatively similar to that of the barium analogue. Further insight into these results is gained from DFT simulations.

**DFT and atomic origin of phonon modes:**

We next performed a series of calculations to determine the character of the phonon modes to corroborate our estimate of the phonon DOS from heat capacity measurements. Since the concentration of oxygen vacancies is relatively low, we considered a stoichiometric $Sr_2CaWO_6$ monoclinic phase, represented with a 40-atom periodic supercell (see SI Figure S40, Table S11). A pseudocubic (α = β = γ = 90°) phase was also considered, details in SI §S14. Phonon DOS was determined by applying Gaussian smearing with 30 $cm^{-1}$ FWHM of the DFT-computed vibrational frequencies.

Figure 6a shows the calculated vibrational DOS overlaid with the phonon density of states predicted by fits to the experimental heat capacity data. The Einstein mode is represented as a Dirac delta, zero everywhere except at the Einstein frequency, and the Debye modes are represented with Equation 3 up to the Debye frequency. s is the oscillator strength and $\theta_D$ is the Debye temperature from Table 1. T is the temperature in Kelvin, $k_B$ is the Boltzmann constant, and ω is the frequency in Hz (note the conversion from wavenumber).

$$D_{Debye}(\omega) = 3s\omega^2 \left(\frac{\hbar}{\theta_D k_B}\right)^3 \tag{3}$$

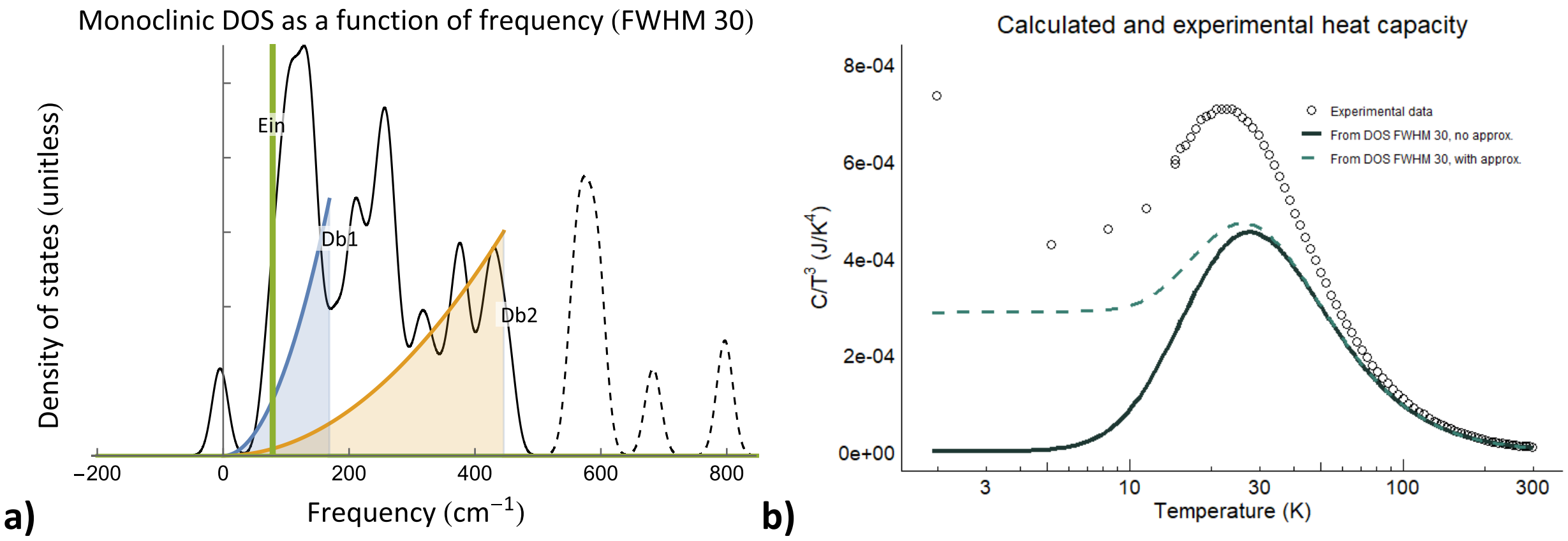

**Figure 6: DFT DOS and calculated heat capacity.** a) Calculated (DFT) vibrational DOS for the monoclinic lattice (black) overlaid with modes fit from experimental heat capacity data (green = Einstein mode with θE = 113 K, blue = Debye mode with θD = 242 K, orange = Debye mode with θD = 641 K). Vibrational DOS as calculated at the gamma point and broadened using Gaussian smearing with FWHM of 30 cm$^{-1}$. Above 450 cm$^{-1}$, a dashed line is used to indicate that these modes are unpopulated at experimental temperatures. b) Heat capacity ($C_v$) computed from DFT DOS shown in a) (lines) and $C_p$ shown in Figure 5a (circles) divided by temperature cubed ($T^3$) from T = 1.9 to 300 K plotted on a logarithmic scale. Note that imaginary (negative frequency) contributions to the DOS were omitted in the integration. Computed data is scaled so that, at high temperature, $C_v$=3nR where n=10 atoms per formula unit and R is the molar gas constant. The solid line depicts $C_v$ computed from the raw DOS shown in a) while the dashed line depicts $C_v$ computed from DOS where a w$^2$ approximation has been applied between the origin and first positive-frequency vibrational frequency, to better approximate the collective low energy phonon modes of a solid.

Figure 6b shows the constant-volume heat capacity calculated from the vibrational density of states shown in Figure 6a overlaid with the experimental constant-pressure heat capacity shown in Figure 5a. Details of the calculation are provided in the Methods section below. At low temperatures, $C_V$ is expected to be approximately equal to $C_p$. The results of different values of FWHM in the Gaussian smearing of calculated vibrational frequencies are also provided in Figure S42. All demonstrate that the calculations capture the key qualitative features of the experimental data, independent of precise choice of broadening, which suggests an accurate DFT representation of the simulated lattice. Importantly, the position of the peak resulting from the Einstein mode is reproduced in the computed data, indicating that the causative behavior – hypothesized from experimental data to be $WO_6$/$CaO_6$ octahedral tilting - is well-captured in this model.

The two Debye modes identified from heat capacity measurements roughly describe the motions of heavy and light atoms, respectively, within the lattice. The higher temperature Debye mode corresponds to the higher energy vibrations of atoms with lower masses and the low energy Debye mode corresponds to the low energy vibrations of atoms with high masses. To establish the precise atomic origin of the phonon modes *via* DFT, we assign them weights corresponding to the magnitude of atomic displacements normalized to unity and summed over atoms of the same type. The results of this analysis are provided in Figure 7 and Figure S41; visualization of the vibrational modes is provided in Movie S1.

Near-zero frequency modes correspond to lattice translations. In the monoclinic lattice, the lowest energy band (60–190 cm$^{-1}$ Figure 6a; phonon mode index 80-117 Figure 7) is dominated by displacements of the Sr and O atoms and contains minor contributions due to Ca and W. This is the only band that includes non-negligible W contributions; hence, it is best described as rotations and displacements of $WO_6$ octahedra coupled with the motion of the framework Sr atoms. This calculated band overlaps with the Einstein and first Debye modes identified *via* experimental heat capacity results, as shown in Figure 6a.

We found that modes located in the interval of 200–450 $cm^{-1}$ (mode index 25-79 in Figure 7) are dominated by Ca and O atoms and the weights of their contributions anticorrelate, as shown in the Figure S41. Accordingly, this band is attributed to symmetric and antisymmetric O-W-O and O-Ca-O bond bending, where Ca participates in the bond-bending motion and W remains stationary. This band overlaps with the second experimentally identified Debye mode, found to have the largest spin-phonon coupling parameter of all the experimentally identified phonon modes.

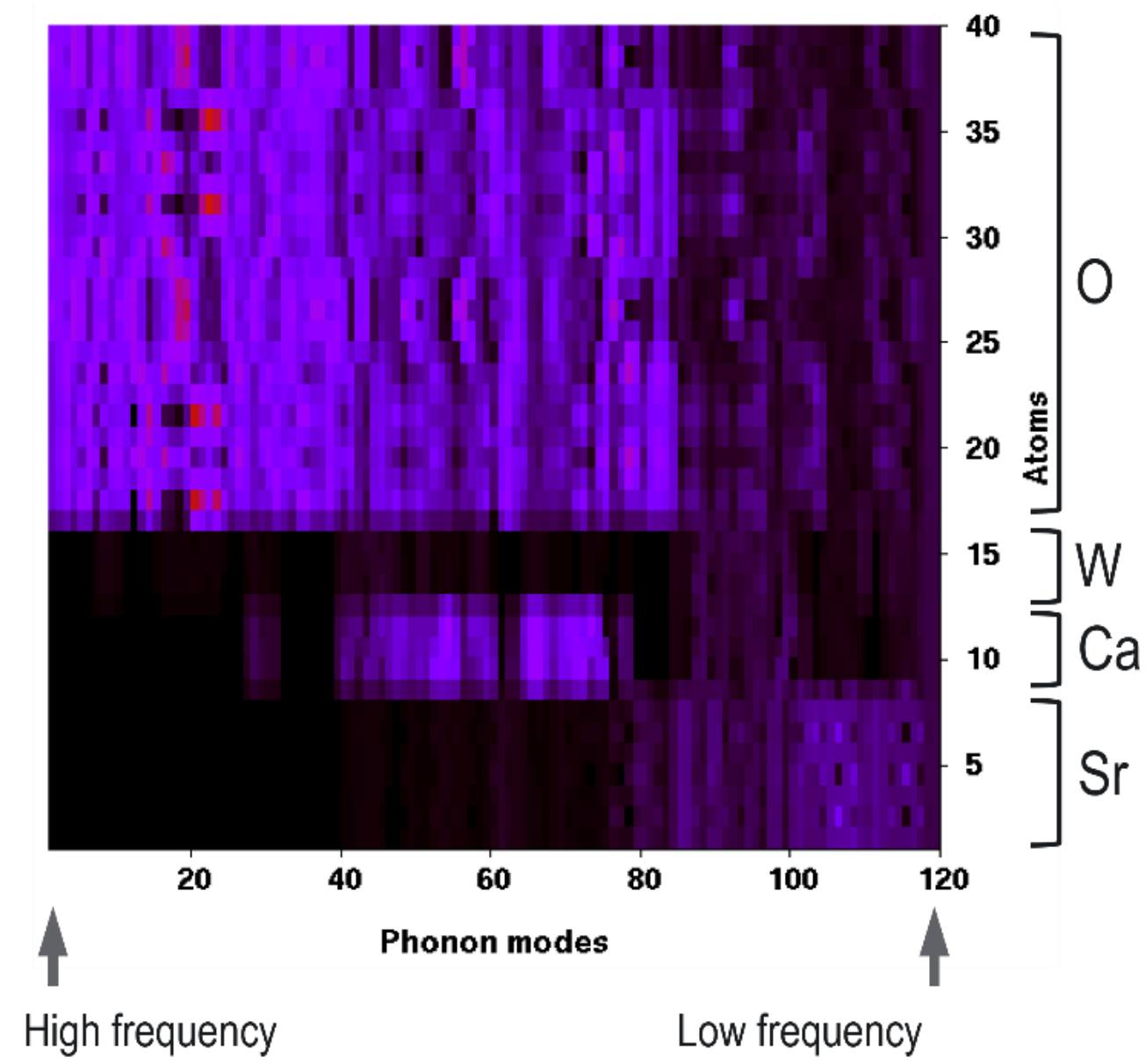

**Figure 7: Heat map representation of the atomic contributions generated by DFT to each phonon mode**. Data is shown for the monoclinic lattice, sorted by index number. Each amplitude vector was normalized to unity. The brighter and darker colors correspond to the larger and smaller amplitude values, respectively.

The high degree of spin-phonon coupling for this mode may be explained by the identification of this mode with O-W-O bending. Displacement of oxygen octahedra, as occurs in lower-frequency bands, has a relatively minor effect on the local electronic environment around the paramagnetic spin centers in $Sr_2CaWO_{6-\delta}$. Conversely, O-W-O bond bending has a strong impact on the symmetry of the electric field experienced by the spin centers. The activation of the second Raman process – linked to this second Debye mode – is correlated with the temperature at which $T_2$ is observed to increase and where EPR and HYSCORE data show increasing spin isotropy. Bending motions increasing the local symmetry of oxygen-deficient tungsten octahedra, or lowering the energy of electron transfer to complete octahedra, may be a source of increasing coherence time.

Finally, high-frequency bands (index 1-24 in Figure 7) are attributed to O-W symmetric and anti-symmetric stretching and include only oxygen motion. These bands, represented with dashed lines in Figure 6a, are not represented in experimental results as they are accessed above the temperature range of our heat capacity and EPR experiments. $Sr_2CaWO_6$ has been shown to undergo a discontinuous structural transition to a cubic structure at T ≅ 1100 K,[53] so above T = 300 K our simulated pseudocubic structure is expected to become more relevant. However, the DOS for both structure types are qualitatively similar in the high-frequency range (Figures S39 and S41).

## Discussion:

Here we have shown the viability of $Sr_2CaWO_{6-\delta}$ as a qubit host by demonstrating millisecond $T_1$ lifetimes at T ~ 10 K. Through HYSCORE and EPR measurements we were able to identify $W^{5+}$ spin centers as the dominant defect centers in these materials with spin localization primarily on oxygen. Decreasing the spin orbit coupling constant by our design strategy led to slightly shorter spin-lattice relaxation times compared to the Ba analogue studied previously; however, the unusual increase in $T_2$ observed between T = 50 and 100 K provides evidence that similar methods could be employed to successfully increase coherence times.

A key component of our original hypothesis was that decreased spin orbit coupling would help to decouple the spin center from lattice effects, but due to the role of the lattice sites (specifically tungsten octahedra) in determining the electronic environment of the paramagnetic spin, this was proved false. The higher g-tensor observed for the paramagnetic electrons in the barium analogue is likely due to higher symmetry of the lattice site[54] and produces longer coherence times at low temperature. The decrease in $T_1$ observed *via* replacement of the A-site cation barium by strontium in this work is limited only to low temperatures and appears related to the increased role of the direct process in spin-lattice relaxation. The overall magnitude of the associated Einstein mode is less than that of the barium analogue, resulting also in a lower heat capacity.

At higher temperatures, a slight reduction in spin-phonon coupling – likely due in part to the replacement of barium with lighter strontium - combined with fewer nuclear spins results in slightly higher $T_1$ values for the new compound. This sets the limit of $T_2$ higher for $Sr_2CaWO_{6-\delta}$ beyond about 50 K. Notably, this is an example of T2 values affected by lattice vibrations even when $T2 \ll T1$ - in contrast to the conventional wisdom - which reinforces the importance of studies on lattice vibration when designing optimized qubit host materials. As demonstrated here, a simple replacement of a cation can affect both $T_1$ and $T_2$ in a variety of interrelated ways even with otherwise minimal changes to the overall lattice.

The peculiar increase in $T_2$ with temperature in the strontium system serves to produce much longer coherence times at higher temperatures despite low-temperature limitations stemming from lower $T_1$. EPR measurements indicate that the spin responsible for these long coherence times occupies a highly symmetrical electronic environment ($g \cong 1.943$, close to free electron $g_e = 2.00232$) which is increasingly activated at higher temperatures while anisotropic spin effects are reduced. The importance of a symmetrical electronic environment on prolonging relaxation times has been established previously for $3d^1$ molecular[54] and $3d^9$ ionic systems,[55] but the thermal activation of a higher-symmetry site created by oxygen vacancies in the dilute limit has not been observed before.

Local symmetry in this system increases as phonon modes are populated. This leads to a reduction of nuclear spin coupling as the W site approaches ideal octahedral symmetry. We propose that a combination of 1) electron motion into complete oxygen octahedra and 2) lattice effects serving to increase the electronic symmetry of incomplete octahedra together produce ever more symmetric environments as temperature increases. Future materials design work should incorporate these strategies to increase the symmetry of spin environments, bringing the electron g tensor closer to that of a free electron, and thus prolonging relaxation times. Finally, as-yet unexplained detrimental aging of the desired phase is also observed in this work and should be more concretely probed prior to application of this system in quantum information technologies.

## Methods

All data were processed using a combination of Xepr, Python 2.7 and 3.11.4, JupyterNotebook version 7.0.8 , Origin Pro 2015, RStudio 2024.04.2 running R version 4.1.2, Mathematica 14.1_WIN, and MATLAB R2018b and R2024a. A sample key is provided in SI §S1.

### Synthesis of $Sr_2CaWO_6$

Polycrystalline $Sr_2CaWO_6$ powder was synthesized using solid state synthesis by reacting loose powders of $SrCO_3$ (Strem Chemicals, 99.99%) $CaCO_3$ (Noah Chemicals, 99.98%) and $WO_3$ (Noah Chemicals, 99.99%) at 650 ˚C, then 1000 ˚C and finally 1250 ˚C for 24 hours per heating in air with a heating and cooling rate of 100 ˚C/hr. The resulting beige polycrystalline powder was compacted into rods using a hydrostatic press at 70 MPa followed by sintering at 1250 ˚C in air for another 24 hours. The resulting rods were yellow in color. In December 2019, these rods were melted at 40% laser power (5 x 200 W GaAs lasers – 976 nm) in a laser diode floating zone (LDFZ) furnace (Crystal Systems Inc FD-FZ-5-200-VPO-PC). A stable floating zone was maintained by counterrotating the rods at 10 rpm and a travelling speed of 10 mm/hr under Ar gas flowing at 2.5 L/min. The obtained crystal was dark blue and ground powder was light blue, differing from the polycrystalline growth rods, indicating the presence of oxygen vacancies in the sample.

The resulting crystal was immediately taken for PXRD. ATR-IR, heat capacity, EPR, and HYSCORE measurements were taken within six months. Approximately four years passed before SCXRD, TGA, and magnetization measurements were performed.

In an attempt to quantify uncertainties associated with natural aging of the sample (see §S9), additional polycrystalline $Sr_2CaWO_6$ powder was synthesized again in 2023 using an optimized solid state synthesis by reacting dried powders of $SrCO_3$, $CaCO_3$, and $WO_3$ for 24 hours at 1000 °C followed by two successive 24 hour heatings at 1250 °C. The material was ground and mixed thoroughly between each heating step. Peaks belonging to $Sr_2WO_5$ were detected using PXRD in samples heated to 1000°C but eliminated with 1250 °C heatings. Intermediate grinding was found to be essential to eliminate $SrWO_4$, and in poorly ground samples an additional heating of 24 hours at 1350 °C or hotter was required to produce single-phase $Sr_2CaWO_6$ (see Figure S28). The resulting single-phase material was analyzed via PXRD, TGA, ATR-IR, and magnetization within the next six months.

### Powder X-ray diffraction

Laboratory-based room temperature x-ray diffraction patterns for phase identification were collected using a Bruker D8 Focus diffractometer with $CuK_\alpha$ radiation in the 15-120˚ 2θ range (Figures S1 and S2). Ground silicon SRM 640d (space group Fd-3m, #227, a = 5.431179 Å)[56] was added to allow accurate determination of lattice parameters (Table S3).

Low temperature PXRD patterns were collected with a Bruker D8 Advance with an Oxford Cryosystems PheniX cryocontroller with $CuK_{\alpha}$ radiation and 6 mm tube and 0.6 mm detector slits. Freshly grown single-crystalline material (Sample SSC-19) was measured using a high-resolution scintillation counter from T = 12-297.6 K (room temperature) in the 26-32° and 53-79° 2θ ranges, referred to as low-angle and high-angle data, respectively. See SI §S2 for discussion. Polycrystalline material grown in 2024 (Sample SPC-24-2) was measured with a Bruker LynxEye detector over the range 17-72° 2θ within 1 week of synthesis. Ground silicon SRM 640d (space group #227, Fd-3m, a = 5.431179 Å)[56] was added to allow accurate determination of lattice parameters (Tables S4, S5, and S6).

Phase identification and unit cell determinations were carried out using the Bruker TOPAS 4.2 software (Bruker AXS).[57]

**Single crystal X-ray diffraction**

SCXRD data for $Sr_2CaWO_{6-\delta}$ approximately four years after synthesis (Sample SSC-23) were collected at 213 K. Data were collected using a SuperNova diffractometer (equipped with an Atlas detector) with Mo Kα radiation (λ = 0.71073 Å) under the program CrysAlisPro (ver. CrysAlisPro 1.171.42.49, Rigaku OD, 2022). The same program was used to refine the cell dimensions and for data reduction. The structure was solved with the program SHELXS-2018/3 and was refined on $F^2$ with SHELXL-2018/3.[58] Analytical numeric absorption correction using a multifaceted crystal model was applied using CrysAlisPro (ver. CrysAlisPro 1.171.42.49, Rigaku OD, 2022). The temperature of the data collection was controlled using the Cryojet system (manufactured by Oxford Instruments). Results in SI §S3.

**Q-Band (~34 GHz) Hyperfine Sublevel Correlation (HYSCORE) Spectroscopy**

Q-band HYSCORE spectroscopy was performed at the Caltech EPR Facility at the California Institute of Technology in Pasadena, CA using a Bruker ELEXSYS E-580 pulse EPR spectrometer equipped with a Bruker D2 Q-band resonator. Temperature control was achieved using a ColdEdge ER 4118HV-CF5-L Flexline Cryogen-Free VT cryostat and an Oxford Instruments Mercury ITC temperature controller.

HYSCORE spectra were acquired using the 4-pulse sequence ($\pi/2 - \tau - \pi/2 - t_1 - \pi - t_2 - \pi/2$ – echo), where $\tau$ is a fixed delay, while $t_1$ and $t_2$ are independently incremented by $\Delta t_1$ and $\Delta t_2$, respectively. The time domain data was baseline-corrected (third-order polynomial) to eliminate the exponential decay in the echo intensity, apodized with a Hamming window function, zero-filled to eight-fold points, and fast Fourier-transformed to yield the 2-dimensional frequency domain.

All HYSCORE spectra were simulated using the EasySpin[59] simulation toolbox (version 5.2.36) with MATLAB 2022b using the following Hamiltonian:

$$\hat{H} = \mu_B \vec{B}_0 g \hat{S} + \mu_N g_N \vec{B}_0 \hat{I} + h \hat{S} \cdot \boldsymbol{A} \cdot \hat{I} \quad (4)$$

In this expression, the first term corresponds to the electron Zeeman interaction term where $\mu_B$ is the Bohr magneton, g is the electron spin g-value matrix with principle components g = [$g_x$ $g_y$ $g_z$], and $\hat{S}$ is the electron spin operator; the second term corresponds to the nuclear Zeeman interaction term where $\mu_N$ is the nuclear magneton, $g_N$ is the characteristic nuclear g-value for each nucleus (e.g. $^{183}W$) and $\hat{I}$ is the nuclear spin operator; the third term corresponds to the electron-nuclear hyperfine term, where $\boldsymbol{A}$ is the hyperfine coupling tensor with principle components $\boldsymbol{A}$ = [$A_x$ $A_y$ $A_z$].

**TGA**

Thermogravimetric analysis was performed with a TA SDT Q600 simultaneous TGA/DSC, weight calibrated before use for 2 °C/min, in order to determine the amount of oxygen vacancies in our samples. Each sample was heated under $O_2$ flowing at 20 mL/min to 150 °C at 10 °C/min and held for one hour to drive off any adsorbed water. The mass after the hold was taken to be the initial sample mass. It was then heated, also under $O_2$, to 1000 °C at 2 °C/min and held for four hours. The mass at the end of this hold was taken to be the final sample mass. Data is shown in SI §S7.

**IR data**

ATR-IR data was collected with a Thermo-Nicolet iN5 with iD5 accessory from 525-5000 cm$^{-1}$. All samples were powders. "As purchased" samples were stored under atmospheric conditions, while all "freshly grown" or heated samples were measured within two days of synthesis or removal from a sealed tube. A combination of literature searches and measurements on various oxide and carbonate precursor materials was used to assign the peaks observed, details in SI §S7.

**Magnetization**

Magnetization as a function of temperature data were collected from T = 2-300 K with an applied field of $\mu_B H$ = 1 T using a Quantum Design MPMS 3 system. Samples were non-oriented LDFZ-grown pieces weighing in total ~15 mg (Sample SSC-23) suspended with plastic wrap in a plastic straw. A Curie-Weiss fit was performed on the magnetic susceptibility χ calculated from these data to approximate the number of magnetic sites. The T = 2-15 K region was selected as it was the most linear portion of the low-temperature data. A $\chi_0$ was determined using a binary search algorithm to make this data as linear as possible, then a least squares fit was performed to determine C and the Curie-Weiss temperature $\theta_{CW}$ according to Equation 5.

$$\frac{1}{\chi - \chi_0} = \frac{T - \theta_{CW}}{C} \tag{5}$$

Since a S = 1/2 site (as from a single unpaired electron) should yield a Curie-Weiss constant C of 0.375 emu K/mol in this temperature region, it was determined that 1.18% of W sites in $Sr_2CaWO_{6-\delta}$ and 0.15% of W sites in $Ba_2CaWO_{6-\delta}$ contained unpaired electrons from $W^{5+}$. See SI §10 for data fit.

$$\text{Oxygen per formula unit} = 6 - 6\left(C \times \frac{8}{3}\right) \tag{6}$$

Magnetization as a function of field data were also collected from -5 to 5 T at 2 K, 20 K, and 50 K on both powder and crystalline samples. Samples were an oriented LDFZ-grown piece (Sample SSC-25) weighing ~53 mg and ground powder of the same sample weighing ~37 mg, suspended with plastic wrap in a plastic straw.

**Reduction and reoxidation of freshly grown samples**

Freshly grown samples were reduced using the following method for comparison to the floating zone sample grown in a reduced state. Approximately 15 mg of $Sr_2CaWO_6$ powder was added to the bottom of a 10” quartz tube. A narrow neck was made in the tube at its midpoint, such that a ¼” cylindrical pellet of Zr metal added to the tube was unable to contact the powder. The tube was backfilled with argon gas and evacuated to ~$10^{-3}$ torr. The tube was inserted into a three-zone furnace and a heating program entered such that both sides of the tube reached their target temperatures at the same time, but the Zr pellet achieved a final temperature ≥150 °C hotter than the powder (Table S9). Samples were characterized with PXRD, TGA, and ATR-IR, results in SI §S11.

**X-band EPR experimental protocol**

X-band EPR spectroscopy was performed on crushed microcrystalline powders contained within a 4 mm OD quartz EPR tube within 6 months of synthesis. Samples were ground from LDFZ-grown $Sr_2CaWO_{6-\delta}$ (Sample SSC-19) confirmed by PXRD to be single-phase. EPR data were obtained at T = 70 K at X-band frequency (~0.3 T, 9.5 GHz) on a Bruker E580 X-band spectrometer at the University of Illinois EPR Lab (Urbana, IL) equipped with a 1 kW TWT amplifier (Applied Systems Engineering). Temperature was controlled using an Oxford Instruments CF935 helium cryostat and an Oxford Instruments ITC503 temperature controller (UIUC). T = 50 K, 60 K, and 70 K data were collected for a shorter measurement time and thus display higher noise.

$T_2$ values as a function of temperature (T = 5-300 K) were determined via a Hahn-echo decay experiment utilizing a $\pi/2 - \tau - \pi - \tau -$ echo sequence. Echo decay as a function of increasing delay time τ was measured and fit to a stretched exponential function (again necessitated by a presumed distribution of domain sizes and electron environments). The time constant associated with that decay is the $T_2$ value. Spin-lattice relaxation times measured at 3500 G were the longest at all temperatures and were also the most persistently measurable with increasing temperature, yielding able to be integrated and manipulated at temperatures as high T = 300 K.

EasySpin v 6.05 running in MATLAB R2024a was used to fit the EPR data presented in Figure 2a.[59] Below T = 50 K, the two peaks in the data were best modelled as a combination of three symmetric g tensors. From T = 50-100 K, only two g tensors were required, one for each peak. Beyond T = 100 K, only one peak is present and only one g tensor was required. When plotted as a function of temperature, the g tensor corresponding to the isotropic low-field peak remains nearly constant at $g \cong 1.943$ (Figure 2b). The g tensors responsible for the anisotropic peak, conversely, show some temperature dependence. Figure S32 shows the total contribution of the isotropic peak to the total signal, as a percentage of total peak area, and reproduces the trend visually observed in Figure 2a.

**Fitting of $T_1^{-1}$ data**

To obtain the contributions of each decoherence process to the spin-relaxation lifetime relative to temperature we fit $1/T_1$ with respect to temperature. Equation 9 was used to fit the data, where $A_{dir}$, $A_{ram1}$, $A_{ram2}$, and $A_{loc}$ are coefficients representing one direct, two Raman, and one local process, respectively. Values are reported in Table 2. $J_8$ is the transport integral describing the joint phonon DOS assumed by the Debye model taking the form described in Equation 10.

$$\frac{1}{T_1} = A_{dir}\mathrm{T} + A_{ram1}\left(\frac{T}{\theta_{\mathrm{D1}}}\right)^9 J_8\left(\frac{\theta_{\mathrm{D1}}}{T}\right) + A_{ram2}\left(\frac{T}{\theta_{\mathrm{D2}}}\right)^9 J_8\left(\frac{\theta_{\mathrm{D2}}}{T}\right) + A_{loc}\frac{\mathrm{e}^{\theta_{\mathrm{E}}/T}}{(\mathrm{e}^{\theta_{\mathrm{E}}/\mathrm{T}} - 1)^2} \quad (9)$$

$$J_8\left(\frac{\theta_D}{T}\right) = \int_0^{\theta_D/T} x^8 \frac{e^x}{(e^x - 1)^2} dx \quad (10)$$

A simple monoexponential function was unable to adequately capture the shape of the curve. Attempting to fit the data using a correction to account for spectral diffusion resulted in unrealistic relaxation times below T = 30 K, representing an inability of the model to account for relaxation driven by spectral diffusion. This has been observed previously in systems in which spectral diffusion happens much faster than spin-lattice relaxation.[60] A stretched monoexponential better captured the curvature. Such stretch factors are typically attributable to a range of relaxation times across the sample,[61] which we here ascribe to the existence of more than one paramagnetic electron environment and the variance in domain sizes inherent to inhomogeneously ground microcrystalline samples.

**Heat capacity measurements**

A Quantum Design Physical Properties Measurement System (PPMS) was used for the heat capacity measurements from T = 1.9 to 300 K at $\mu_o H$ = 0 T using the semi-adiabatic method. The data were fit using the equations:

$$\mathrm{C}_{\mathrm{Einstein}} = 3sR\left(\frac{\theta_E}{T}\right)^2 \times \frac{e^{\theta_E/T}}{(e^{\theta_E/T} - 1)^2} \quad (11)$$

$$\mathrm{C}_{\mathrm{Debye}} = 9sR\left(\frac{T}{\theta_D}\right)^3 \times \int_0^{\theta_D/T} \frac{\left(\frac{\theta}{T}\right)^4 e^{\left(\frac{\theta}{T}\right)}}{\left[e^{\left(\frac{\theta}{T}\right)} - 1\right]^2} d\frac{\theta}{T} \quad (12)$$

yielding the fit shown in Figure 5a. The parameters used in this fit, including the Debye ($\theta_D$) and Einstein temperatures ($\theta_E$), are shown in Table 1. R is the molar gas constant and s is the number of oscillators per formula unit.

**DFT**

DFT calculations were performed using the VASP package[62-64] and the PBEsol density functional.[64] The projector-augmented wave potentials were used to approximate the effect of the core electrons.[65] Gamma-centered k-mesh was varied from 2×2×2 to 4×4×4 to ensure the convergence of the structural parameters; 3×3×3 mesh was used for the calculations of vibrational frequencies. The plane-wave basis-set cutoff was set to 600 eV. The total energy was converged to within $10^{-8}$ eV. The vibrational frequencies and phonon modes at the Gamma point were calculated using the finite differences approach and the displacement magnitude of 0.01 Å.

Vibrational DOS was calculated by convoluting vibrational frequencies with Gaussian functions with the full width at half maximum (FWHM) of 30 $cm^{-1}$.

To account for self-interaction of electrons in our DFT simulations, ionization energy and sphericity were estimated for a range of U values. For our d-orbital spin density estimates, an extra electron (fixed doublet state) was added and treated explicitly; a homogeneous background charge was also added to maintain charge neutrality. For our estimates of vacancy spin density, a triplet state was fixed. Additional discussion is provided in SI §S13-S15.

**Calculation of constant volume heat capacity from DFT DOS results**

Phonons were calculated using the finite displacement method of a 3x3x3 supercell. This is too coarse a mesh for a full Brillouin zone integration to obtain the true phonon DOS. For calculation of the heat capacity from the phonon DOS, we first approximated the DOS by applying a Gaussian smearing of 30 $cm^{-1}$ to each calculated mode to capture the effects of finite bandwidth, as detailed in §S15 and pictured in Figure 6a.

To calculate the constant volume heat capacity expected from this DOS, we next omitted all 0 energy acoustic phonons occurring at negative frequencies. Heat capacity was then calculated from the vibrational density of states using the standard equation for the heat capacity of bosons from statistical thermodynamics:[66]

$$C_V = \int \left(\frac{\hbar\nu}{Tk_B}\right)^2 D(\nu) \frac{k_B e^{\frac{\hbar\nu}{Tk_B}}}{\left(e^{\frac{\hbar\nu}{Tk_B}} - 1\right)^2} d\nu \tag{13}$$

where T is temperature in Kelvin, ħ is the reduced Planck constant, ν is the frequency in Hz, and $k_B$ is the Boltzmann constant. D(ν) is the density of states as calculated by DFT and shown in Figure 6a. Numerical integration was done using the trapezoidal Riemann approximation method in R version 4.1.2. The resulting value was normalized such that the heat capacity Cv at high temperature (where a plateau is observed, here T = 1000 K) satisfies the Dulong-Petit Law, Equation 14.

$$\lim_{T\to\infty} C_v = 3nR \tag{14}$$

R is the molar gas constant. Here n = 10 atoms per formula unit.

The resulting data is shown in the solid line of Figure 6b. Lastly, we repeated the calculations, adding a $w^2$ dependence from 0 $cm^{-1}$ to the first finite frequency mode to properly account for the collective low energy phonon modes of a solid. These results are shown in Figure 6b as the dashed line. The calculations were repeated for additional Gaussian broadenings, Figure S42, as confirmation that the calculations capture the key qualitative features of the experimental data independent of broadening.

## Acknowledgements

This work was funded by the U.S. Department of Energy (DOE), Office of Science (SC), National Quantum Information Science Research Centers, Co-Design Center for Quantum

Advantage (C2QA) under contract number DE-SC0012704. Computational modeling at the Pacific Northwest National Laboratory (PNNL) was supported by C2QA (BES, PNNL FWP 76274). This research also used resources of the National Energy Research Scientific Computing Center, a DOE SC User Facility supported by the SC of the U.S. DOE under Contract No. DE-AC02-05CH11231 using NERSC award BES-ERCAP0028497. Use was made of the synthesis facilities of the Platform for the Accelerated Realization, Analysis, and Discovery of Interface Materials (PARADIM), which are supported by the National Science Foundation under Cooperative Agreement No. DMR-2039380. The MPMS3 system used for magnetic characterization was funded by the National Science Foundation, Division of Materials Research, Major Research Instrumentation Program, under Award No. 1828490. EPR studies were performed at the University of Illinois School of Chemical Sciences EPR lab (Urbana, IL) with assistance from Dr. Toby Woods. The Caltech EPR facility acknowledges the Beckman Institute and Dow Next Generation Educator Fund for financial support. TJP gratefully acknowledges the support of an NSF Graduate Research Fellowship (Grant No. DGE-1324585). Figure 1a was created using VESTA.[67] TMM acknowledges discussions with Nathalie de Leon.

## Data availability

Raw data associated with the syntheses and material characterizations in this work are accessible at data.paradim.org. Additional data sets generated during the current study are available from the corresponding author on reasonable request.

## Code availability

The underlying code for this study is not publicly available but may be made available to qualified researchers on reasonable request to the corresponding author.

## Author contributions

MKS and WAP developed and performed the single crystal growth methods, collected and analyzed all heat capacity data, and collected and analyzed all XRD data prior to 2019. MKS developed and performed polycrystalline growth methods and collected ATR-IR data for samples prior to 2019. SMB developed and performed polycrystalline growth methods and collected and analyzed ATR-IR and XRD data for samples after 2019. PVS designed and performed ab initio simulations and analyzed the results. PHO collected and analyzed Q-band HYSCORE data. TJP collected and analyzed EPR data. MAS collected and analyzed SCXRD data. ANN collected magnetization data. SMB applied the cubicity metric, performed reductions, collected TGA data, fit EPR data, and analyzed magnetization data. All authors reviewed and discussed the final manuscript.

## Competing interests

All authors declare no financial or non-financial competing interests.

## References

1. Nielsen, M. A., Chuang, I. & Grover, L. K., Quantum Computation and Quantum Information. *Am. J. Phys.* **70,** 4; 10.1119/1.1463744 (2002).
2. Feynman, R. P., Simulating Physics with Computers. *Int. J. Theor. Phys.* **21,** 467-488; 10.1007/bf02650179 (1982).
3. Degen, C. L., Reinhard, F. & Cappellaro, P., Quantum Sensing. *Rev. Mod. Phys.* **89,** 035002; 10.1103/revmodphys.89.035002 (2017).
4. Awschalom, D. *et al.*, Development of Quantum Interconnects (QuICs) for Next-Generation Information Technologies. *PRX Quantum* **2,** 017002; 10.1103/prxquantum.2.017002 (2021).
5. National Academies of Sciences, Engineering, and Medicine *Advancing Chemistry and Quantum Information Science: An Assessment of Research Opportunities at the Interface of Chemistry and Quantum Information Science in the United States* (The National Academies Press, 2023).
6. Bedow, J., Mascot, E., Hodge, T., Rachel, S. & Morr, D. K., Simulating topological quantum gates in two-dimensional magnet-superconductor hybrid structures. *npj Quantum Mater.* **9,** 99; 10.1038/s41535-024-00703-w (2024).
7. Han, J.-H. *et al.*, Weak-coupling to strong-coupling quantum criticality crossover in a Kitaev quantum spin liquid α-RuCl3. *npj Quantum Mater.* **8,** 33; 10.1038/s41535-023-00563-w (2023).
8. Brown, K. R., Chiaverini, J., Sage, J. M. & Häffner, H., Materials challenges for trapped-ion quantum computers. *Nat. Rev. Mater.* **6,** 892–905; 10.1038/s41578-021-00292-1 (2021).
9. Bruzewicz, C. D., Chiaverini, J., McConnell, R. & Sage, J. M., Trapped-Ion Quantum Computing: Progress and Challenges. *Appl. Phys. Rev.* **6,** 021314; 10.1063/1.5088164 (2019).
10. Devoret, M. H. & Schoelkopf, R. J., Superconducting Circuits for Quantum Information: An Outlook. *Science* **339,** 1169-1174; 10.1126/science.1231930 (2013).
11. Gaita-Ariño, A., Luis, F., Hill, S. & Coronado, E., Molecular spins for quantum computation. *Nat. Chem.* **11,** 301-309; 10.1038/s41557-019-0232-y (2019).
12. Calafell, I. A. *et al.*, Quantum computing with graphene plasmons. *npj Quantum Inf.* **5,** 37; 10.1038/s41534-019-0150-2 (2019).
13. Duan, L.-M., Lukin, M., Cirac, I. & Zoller, P., Long-Distance Quantum Communication with Atomic Ensembles and Linear Optics. *Nature* **414,** 413-418; 10.1038/35106500 (2001).
14. Bockstedte, M., Schütz, F., Garratt, T., Ivády, V. & Gali, A., Ab initio description of highly correlated states in defects for realizing quantum bits. *npj Quant. Mater.* **3,** 31; 10.1038/s41535-018-0103-6 (2018).
15. van der Laan, K. J., Hasani, M., Zheng, T. & Schirhagl, R., Nanodiamonds for In Vivo Applications. *Small* **14,** 1703838; 10.1002/smll.201703838 (2018).
16. Rondin, L. *et al.*, Magnetometry with Nitrogen-Vacancy Defects in Diamond. *Rep. Prog. Phys.* **77,** 056503; 10.1088/0034-4885/77/5/056503 (2014).

17. Yu, C.-J., von Kugelgen, S., Laorenza, D. W. & Freedman, D. E., A Molecular Approach to Quantum Sensing. *ACS Cent. Sci.* **7,** 712-723; 10.1021/acscentsci.0c00737 (2021).
18. Leuenberger, M. N. & Loss, D., Quantum Computing in Molecular Magnets. *Nature* **410,** 789–793; 10.1038/35071024 (2001).
19. Pla, J. J. *et al.*, A Single-Atom Electron Spin Qubit in Silicon. *Nature* **489,** 541–545; 10.1038/nature11449 (2012).
20. Morita, Y., Suzuki, S., Sato, K. & Takui, T., Synthetic Organic Spin Chemistry for Structurally Well-Defined Open-Shell Graphene Fragments. *Nature Chem* **3,** 197-204; 10.1038/nchem.985 (2011).
21. Zadrozny, J. M., Niklas, J., Poluektov, O. G. & Freedman, D. E., Millisecond Coherence Time in a Tunable Molecular Electronic Spin Qubit. *ACS Cent. Sci.* **1,** 488-492; 10.1021/acscentsci.5b00338 (2015).
22. Stamp, P. C. E. & Gaita-Ariño, A., Spin-Based Quantum Computers Made by Chemistry: Hows and Whys. *Journal of Materials Chemistry* **19,** 1718-1730; 10.1039/b811778k (2009).
23. Nellutla, S., Morley, G. W., van Trol, J., Pati, M. & Dalal, N. S., Electron Spin Relaxation and K39 Pulsed ENDOR Studies on Cr5+-Doped K3NbO8 at 9.7 and 240 GHz. *Phys. Rev. B* **78,** 054426; 10.1103/physrevb.78.054426 (2008).
24. Yu, C.-J. *et al.*, Long Coherence Times in Nuclear Spin-Free Vanadyl Qubits. *J. Am. Chem. Soc.* **138,** 14678-14685; 10.1021/jacs.6b08467 (2016).
25. Yamamoto, T. *et al.*, Extending spin coherence times of diamond qubits by high-temperature annealing. *Phys. Rev. B.* **88,** 075206; 10.1103/physrevb.88.075206 (2013).
26. Balasubramanian, G. *et al.*, Ultralong spin coherence time in isotopically. *Nature Materials* **8,** 383-387; 10.1038/nmat2420 (2009).
27. Bader, K. *et al.*, Room temperature quantum coherence in a potential molecular qubit. *Nature Comm.* **5,** 5304; 10.1038/ncomms6304 (2014).
28. Bulancea-Lindvall, O., Eiles, M. T., Son, N. T., Abrikosov, I. A. & Ivády, V., Isotope-Purification-Induced Reduction of Spin-Relaxation and Spin-Coherence Times in Semiconductors. *Phys. Rev. App.* **19,** 064046; 10.1103/physrevapplied.19.064046 (2023).
29. Sinha, M. *et al.*, Introduction of Spin Centers in Single Crystals of Ba2CaWO6−δ. *Phys. Rev. Materials* **3,** 125002; 10.1103/physrevmaterials.3.125002 (2019).
30. Mirzoyan, R., Kazmierczak, N. P. & Hadt, R. G., Deconvolving Contributions to Decoherence in Molecular Electron Spin Qubits: A Dynamic Ligand Field Approach. *Chem.– Eur. J.* **27,** 9482-9494; 10.1002/chem.202100845 (2021).
31. Atzori, M. *et al.*, Quantum Coherence Times Enhancement in Vanadium(IV)-Based Potential Molecular Qubits: The Key Role of the Vanadyl Moiety. *J. Am. Chem. Soc.* **138,** 11234-11244; 10.1021/jacs.6b05574 (2016).
32. Lunghi, A. & Sanvito, S., Multiple Spin–Phonon Relaxation Pathways in a Kramer Single-Ion Magnet. *J. Chem. Phys.* **153,** 174113; 10.1063/5.0017118 (2020).

33. Escalera-Moreno, L., Baldoví, J. J., Gaita-Ariño, A. & Coronado, E., Spin States, Vibrations and Spin Relaxation in Molecular Nanomagnets and Spin Qubits: A Critical Perspective. *Chem. Sci.* **9,** 3265-3275; 10.1039/c7sc05464e (2018).
34. Pearson, T. J., Laorenza, D. W., Krzyaniak, M. D., Wasielewski, M. R. & Freedman, D. E., Octacyanometallate Qubit Candidates. *Dalton Trans.* **47,** 11744-11748; 10.1039/c8dt02312c (2018).
35. Jackson, C. E., Moseley, I. P., Martinez, R., Sung, S. & Zadrozny, J. M., A reaction-coordinate perspective of magnetic relaxation. *Chem. Soc. Rev.* **50,** 6684-6699; 10.1039/d1cs00001b (2021).
36. Berliner, L. J., Eaton, G. R. & Eaton, S. S. eds. *Distance Measurements in Biological Systems by EPR* (Springer, 2000).
37. Albino, A. *et al.*, First-Principles Investigation of Spin–Phonon Coupling in Vanadium-Based Molecular Spin Quantum Bits. *Inorg. Chem.* **22,** 11249-11265; 10.1039/d0cp00852d (2019).
38. Paddison, J. A. M. *et al.*, Cubic double perovskites host noncoplanar spin textures. *npj Quantum Mater.* **9,** 48; 10.1038/s41535-024-00650-6 (2024).
39. Bhui, A. *et al.*, Intrinsically Low Thermal Conductivity in the N-Type Vacancy-Ordered Double Perovskite Cs2SnI6: Octahedral Rotation and Anharmonic Rattling. *Chem. Mater.* **34,** 3301-3310; 10.1021/acs.chemmater.2c00084 (2022).
40. Chen, H., Magnetically driven orbital-selective insulator–metal transition in double perovskite oxides. *npj Quant. Mater.* **3,** 57; 10.1038/s41535-018-0131-2 (2018).
41. Gordon, L. *et al.*, Quantum computing with defects. *MRS Bulletin* **38,** 802-807; 10.1557/mrs.2013.206 (2013).
42. Kazmierczak, N. P., Mirzoyan, R. & Hadt, R. G., The Impact of Ligand Field Symmetry on Molecular Qubit Coherence. *J. Am. Chem. Soc.* **143,** 17305-17315; 10.1021/jacs.1c04605 (2021).
43. Chirolli, L. & Burkard, G., Decoherence in solid-state qubits. *Adv. in Phys.* **57,** 225-285; 10.1080/00018730802218067 (2008).
44. Amassah, G., Mitchell, D. G., Hovey, T. A., Eaton, S. S. & Eaton, G. R., Electron Spin Relaxation of SO2− and SO3− Radicals in Solid Na2S2O4, Na2S2O5, and K2S2O5. *Appl. Magn. Reson.* **54,** 849-867; 10.1007/s00723-023-01569-0 (2023).
45. Ngendahimana, T., Moore, W., Canny, A., Eaton, S. S. & Eaton, G. R., Electron Spin Relaxation Rates of Radicals in Irradiated Boron Oxides. *Appl. Magn. Reson.* **54,** 359-370; 10.1007/s00723-022-01514-7 (2023).
46. Lunghi, A. & Sanvito, S., How do phonons relax molecular spins? *Sci. Adv.* **5,** eaax7163; 10.1126/sciadv.aax7163 (2019).
47. Yu, C. *et al.*, Spin and Phonon Design in Modular Arrays of Molecular Qubits. *Chemistry of Materials* **32,** 10200-10206; 10.1021/acs.chemmater.0c03718 (2020).
48. Lunghi, A., Totti, F., Sanvito, S. & Sessoli, R., Intra-molecular origin of the spin-phonon coupling in slow-relaxing molecular magnets. *Chem. Sci.* **8,** 6051-6059; 10.1039/c7sc02832f (2017).

49. Šimėnas, M., Ciupa, A., Mączka, M., Pöppl, A. & Banys, J., EPR Study of Structural Phase Transition in Manganese-Doped [(CH3)2NH2][Zn(HCOO)3] Metal–Organic Framework. *J. Phys. Chem. C* **119,** 24522-24528; 10.1021/acs.jpcc.5b08680 (2015).
50. Ramirez, A. P. & Kowach, G. R., Large Low Temperature Specific Heat in the Negative Thermal Expansion Compound ZrW2O8. *Phys. Rev. Lett.* **80,** 4903; 10.1103/physrevlett.80.4903 (1998).
51. Dutta, M. *et al.*, Ultralow Thermal Conductivity in Chain-like TlSe Due to Inherent Tl+ Rattling. *J. Am. Chem. Soc.* **141,** 20293-20299; 10.1021/jacs.9b10551 (2019).
52. Hoffman, S. K. & Lejewski, S., Phonon spectrum, electron spin–lattice relaxation and spin–phonon coupling of Cu2+ ions in BaF2 crystal. *Journal of Magnetic Resonance* **252,** 49-54; 10.1016/j.jmr.2014.12.015 (2015).
53. Gateshki, M. & Igartua, J. M., Crystal structures and phase transitions of the double-perovskite oxides Sr2CaWO6 and Sr2MgWO6. *J. Phys.: Condens. Matter* **16,** 6639-6649; 10.1088/0953-8984/16/37/001 (2004).
54. Martens, M., Franco, G., Dalal, N. S., Bertaina, S. & Chiorescu, I., Spin-orbit coupling fluctuations as a mechanism of spin decoherence. *Phys. Rev. B* **96,** 180408(R); 10.1103/physrevb.96.180408 (2017).
55. Kinyon, J. S. *et al.*, High-frequency EPR study of the unusual multiferroic NH4CuCl3. *Polyhedron* **177,** 114255; 10.1016/j.poly.2019.114255 (2020).
56. SRM 604d; *Silicon*; National Institute of Standards and Technology; U.S. Department of Commerce: Gaithersburg, MD (9 July 2009).
57. Coelho, A. A., TOPAS and TOPAS-Academic: an optimization program integrating computer algebra and crystallographic objects written in C++. *J. Appl. Cryst.* **51,** 210-218; 10.1107/s1600576718000183 (2018).
58. Sheldrick, G. M., Crystal structure refinement with SHELXL. *Acta Crys. C.* **71,** 3-8; 10.1107/s2053229614024218 (2015).
59. Stoll, S. & Schweiger, A., EasySpin, a comprehensive software package for spectral simulation and analysis in EPR. *J. Magn. Reson.* **178,** 42-55; 10.1016/j.jmr.2005.08.013 (2006).
60. Kevan, L. & Schwartz, R. N. eds. *Time Domain Electron Spin Resonance* (Wiley, 1979).
61. Šimėnas, M. *et al.*, EPR of Structural Phase Transition in Manganese- and Copper-Doped Formate Framework of [NH3(CH2)4NH3][Zn(HCOO)3]2. *J. Phys. Chem. C* **120,** 19751–19758; 10.1021/acs.jpcc.6b07389 (2016).
62. Kresse, G. & Furthmüller, J., Efficient iterative schemes for ab initio total-energy calculations using a plane-wave basis set. *Phys. Rev. B* **54,** 11169; 10.1103/physrevb.54.11169 (1996).
63. Kresse, G. & Joubert, D., From ultrasoft pseudopotentials to the projector augmented-wave method. *Phys. Rev. B.* **59,** 1758−1775; 10.1103/physrevb.59.1758 (1999).
64. Perdew, J. P. *et al.*, Restoring the Density-Gradient Expansion for Exchange in Solids and Surface. *Phys. Rev. Lett.* **100,** 136406; 10.1103/physrevlett.100.136406 (2008).
65. Blöchl, P. E., Projector augmented-wave method. *Phys. Rev. B.* **50,** 17953; 10.1103/physrevb.50.17953 (1994).

66. Grimvall, G. *Thermophysical properties of materials* (Elsevier Science & Technology, 1999).
67. Momma, K. & Izumi, F., VESTA 3 for three-dimensional visualization of crystal, volumetric and morphology data. *J. Appl. Cryst.* **44,** 1272-1276; 10.1107/s0021889811038970 (2011).